\documentclass[sn-mathphys-num]{sn-jnl}

\usepackage{graphicx}%
\usepackage{multirow}%
\usepackage{amsmath,amssymb,amsfonts}%
\usepackage{amsthm}%
\usepackage{mathrsfs}%
\usepackage[title]{appendix}%
\usepackage{xcolor}%
\usepackage{textcomp}%
\usepackage{manyfoot}%
\usepackage{booktabs}%
\usepackage{algorithm}%
\usepackage{algorithmicx}%
\usepackage{algpseudocode}%
\usepackage{listings}%
\usepackage{verbatim}
\usepackage{enumitem}
 \usepackage{comment}

\begin{document}

\title[Article Title]{Finite Population Dynamics Resolve the Central Paradox of the Inspection Game}


\author[1]{\fnm{Bianca Y.  S.} \sur{Ishikawa}}

\author[1]{\fnm{Jos\'e F.} \sur{Fontanari}}

\affil[1]{\orgdiv{Instituto de F\'{\i}sica de S\~ao Carlos}, \orgname{Universidade de S\~ao Paulo}, \city{S\~ao Carlos}, \postcode{ 13566-590}, \state{S\~ao Paulo}, \country{Brazil}}



\abstract{The Inspection Game is the canonical model for the strategic conflict between law enforcement (inspectors) and citizens (potential criminals). Its classical Mixed-Strategy Nash Equilibrium (MSNE) is afflicted by a paradox: the equilibrium crime rate is independent of both the penalty size ($p$) and the crime gain ($g$), undermining the efficacy of deterrence policy.  We re-examine this challenge using evolutionary game theory,  focusing on the long-term fixation probabilities of strategies in finite, asymmetric population sizes subject to demographic noise. The deterministic limit of our model exhibits stable limit cycles  around  the MSNE, which coincides with the neutral fixed point of the equilibrium analysis.  Crucially, in finite populations, demographic noise drives the system away from this cycle and toward absorbing states. Our results demonstrate that high absolute penalties $p$ are highly effective at suppressing crime by influencing the geometry of the deterministic dynamics, which in turn biases the fixation probability toward the criminal extinction absorbing state, thereby restoring the intuitive role of $p$. Furthermore, we reveal a U-shaped policy landscape where both high penalties and light penalties (where $p \approx g$) are successful suppressors, maximizing criminal risk at intermediate penalty levels. Most critically, we analyze the realistic asymptotic limit of extreme population sizes  asymmetry, where inspectors are exceedingly rare. In this limit, the system's dynamic outcome is entirely decoupled from the citizen payoff parameters $p$ and $g$, and is instead determined by the initial frequency of crime relative to the deterrence threshold (the ratio of inspection cost to reward for catching a criminal). This highlights that effective crime suppression  requires managing the interaction between deterministic dynamics, demographic noise, and initial conditions.

}


\maketitle

\section{Introduction}\label{sec:1}

The problem of law enforcement and deterrence remains central to the stability and economic prosperity of societies, forming a cornerstone of socio-economic analysis that dates back to Becker's  seminal work on the economics of crime \cite{Becker_1974}. The interaction at the heart of this challenge---citizens deciding whether to commit a crime versus authorities deciding whether to inspect---is classically modeled as the Inspection Game. This game was famously introduced by Tsebelis to exemplify the Robinson Crusoe Fallacy: the critical error of mistaking a strategic situation for a predictable problem of individual decision-making (e.g., treating speeding as a decision-theoretic problem like ``What is the probability of being caught?") rather than a dynamic game against an adversary whose actions are conditional on your own \cite{Tsebelis_1989}.  Analyzing the long-term, collective outcomes of such dynamic conflicts requires moving beyond the static equilibrium predictions of classical game theory and embracing the adaptive dynamics of evolutionary game theory \cite{Maynard_1982,Hofbauer_1998}.

Evolutionary game theory provides a powerful framework for studying how population-level strategies evolve based on their relative success, proving crucial for understanding a wide array of fundamental social dilemmas, including the evolution of cooperation \cite{Hamilton_1975, Axelrod_1984, Perc_2017, Wang_2023}, the dynamics of corruption \cite{Katsikas_2016,Aga_2024,Marino_2025}, and, highly relevant in the modern digital landscape, the evolution of lying and deception \cite{Sober_1994,Capraro _2019,Fontanari_2023,Vieira_2025}---a phenomenon intrinsically related to the propagation of disinformation. In this paper, we employ evolutionary game theory  and finite population simulations to re-examine the Inspection Game, focusing on the counter-intuitive dynamics of crime, enforcement, and the ultimate fate of crime in a realistic environment subject to demographic noise. 

The core motivation for this study lies in a critical shortcoming of the Inspection Game's classical analysis. The game possesses a unique Mixed-Strategy Nash Equilibrium (MSNE), the solution of which yields a profound, counter-intuitive result often referred to as the central paradox of law enforcement \cite{Rauhut_2009,Rauhut_2015,Rauhut_2021}: the equilibrium frequency of criminal behavior is determined only by the enforcement cost parameters, while being independent of the crime gain ($g$) and the penalty size ($p$). This finding implies that increasing the penalty for a crime will not, in the long run, reduce the rate of criminal activity, thereby severely challenging the policy relevance of the static model.  To move beyond the limitations of the MSNE, the evolutionary game theory  framework has been expanded to explore more nuanced aspects of law enforcement: models incorporating ordinary non-punishing citizens and spatial interactions demonstrate that crime is often recurrent and is a system-immanent collective behavior, reinforcing the need to move beyond static analysis \cite{Perc_2013}. Additionally, work exploring heterogeneous punishment strategies has shown that diversity in payoff parameters drastically increases the system's complexity and highlights both the benefits and pitfalls of enforcement strategies \cite{Perc_2015}.  Despite these advances, a full resolution of the central paradox—demonstrating the absolute effect of penalty magnitude on the long-term fixation of strategies—remains elusive within deterministic and infinite-population frameworks.  To address this critical failure of the deterministic approach, we analyze the Inspection Game dynamics in a finite population context, introducing demographic noise. This stochastic approach allows us to move beyond stable oscillations and calculate the long-term fixation probabilities of criminal behavior, revealing a stochastic mechanism through which parameters like the penalty size $p$ can, in fact, restore their intuitive role as effective suppressors of crime.

Our analysis reveals several new and surprising findings that fundamentally reshape the understanding of law enforcement dynamics. First, contrary to the deterministic MSNE prediction that criminal frequency is independent of citizens' payoff parameters, we demonstrate that high absolute penalty $p$ is highly effective at suppressing crime in finite populations. This resolution of the classical paradox stems from demographic noise driving the system to the criminal extinction absorbing state when the penalty is large (the high-penalty regime). Second, and equally counter-intuitive, we find that a light penalty (the regime where the penalty $p$ is similar in magnitude to the crime gain $g$) also robustly suppresses crime, establishing a U-shaped policy landscape where the risk of criminal dominance is maximized at moderate penalties. Third, we analyze the asymptotic limit of extreme population size asymmetry,  where the inspector population is much smaller than the citizen population. Crucially, in this highly resource-constrained regime, the general deterrent effect of $p$ vanishes, and the long-term fate of crime is instead entirely decoupled from the citizen payoff parameters $p$ and $g$. Instead, it depends on  the value of  the initial criminal frequency ($x_0$) relative to the deterrence threshold—the ratio between the inspection cost ($k$) and the reward for a successful catch ($r$). The most paradoxical result here is the observed dominance of the initially rare strategy: if $x_0$ is below this threshold (i.e., crime is initially rare), the system is driven toward a population of criminals only. Conversely, if $x_0$ is above the threshold, criminal extinction occurs. These stochastic and non-linear effects highlight that crime suppression policy success relies less on average deterministic behavior and more on managing the system's susceptibility to demographic noise and initial conditions.

The remainder of this paper is structured as follows. We begin in Sec. \ref{sec:ig} by formally introducing the Inspection Game, defining its payoff matrices, and revisiting the classical  MSNE analysis that yields the central paradox of deterrence. Next, in Sec. \ref{sec:id}, we introduce the imitation dynamics, which serves as the stochastic algorithm for simulating the finite population version of the game. We then use the birth-death process framework in Sec. \ref{sec:bd} to justify and derive the replicator equations in the deterministic limit for populations of different sizes, a crucial step given the natural asymmetry between the inspector and citizen population sizes. The full expressions for the underlying transition probabilities are presented in Appendix \ref{appA}.  Sec. \ref{sec:re} then analyzes the solutions of these replicator equations, focusing on a phase plane analysis to characterize the stable oscillations and their amplitude. In Sec. \ref{sec:fp}, we present the core results from the Monte Carlo simulations of the finite population imitation dynamics, where we demonstrate the resolution of the MSNE paradox and the emergence of the paradoxical fixation outcomes. Finally, Sec. \ref{sec:disc} provides a summary of our findings and concluding remarks.

\section{The Inspection  Game}\label{sec:ig}

Here we use the notation of the inspection game studied by Rauhut \cite {Rauhut_2015}. There are two different groups of actors, where members of one group can decide to commit a crime or not and members of the other to inspect or not. The first group are called citizens and the second group inspectors. We assume the number of citizens and inspectors are $N$ and $M$, respectively. The following payoff matrices describe the utility for a single citizen and a single inspector interaction.

Citizens  earn $g$
for the crime, but face punishment costs $p$ if caught.  If citizens commit no crime,  their payoff is zero.  The payoff matrix for citizens  is
\begin{equation}\label{citizen}
\begin{array}{lcc}
& \mbox{Inspect} & \mbox{Not Inspect} \\
 \mbox{Crime} & g-p & g \\
\mbox{No Crime}& 0 & 0
\end{array} 
\end{equation}
where  $g$ and $p$ are non-negative parameters.   We assume undetected crime is profitable ($g>0$) and that punishment costs must be higher than profits from crime ($p>g$), which ensures that $g-p<0$. This means that payoffs are higher for not committing crimes than to commit a crime and receive a punishment for sure.

Inspectors  can invest inspection costs $k$ to detect the action of the citizen and earn the reward $r$ for a successful detection of a crime.  No inspection yields the baseline payoff of zero.  This is summarized in the payoff matrix for inspectors
\begin{equation}\label{inspector}
\begin{array}{lcc}
 & \mbox{Crime} & \mbox{No Crime} \\
\mbox{Inspect} & r-k& -k \\
\mbox{Not Inspect}& 0 & 0 
\end{array} 
\end{equation}
where,  as before,  $r$ and $k$ are non-negative parameters. 
We assume that the reward for a successful detection of a crime must be larger than the cost of inspection ($r>k$).

The citizen's best response is to commit a crime if they are not inspected and not to commit a crime if they are inspected. The inspector's best response is to inspect if the citizen is criminal and not to invest inspection costs if there is no crime. This configuration of cyclical best responses (a non-coordination game) means that there is no Nash Equilibrium in pure strategies. The solution lies in the unique Mixed-Strategy Nash Equilibrium (MSNE), where both actors randomize their actions \cite{Rauhut_2015}.

Let $x$ be the probability that the citizen chooses Crime, and $y$ be the probability that the inspector chooses Inspect. In the MSNE, each player must choose a probability that makes the opponent indifferent between their two pure strategies:
\begin{itemize}
\item[-] The citizen must choose $x^*$ such that the inspector is indifferent between Inspect and Not Inspect.  This requires that the expected payoff for the inspector when  choosing  the pure strategy Inspect, $ x (r-k) - (1-x) k = xr-k$,  equals the expected payoff of  the pure strategy Not Inspect,  which is  $0$.  Setting $xr-k=0$  yields the equilibrium crime rate $x^* = k/r$. 
\item[-] The inspector must choose $y^*$  such that the citizen is indifferent between Crime and No crime. This requires that the expected payoff of the pure strategy Crime,  $y (g-p) + (1-y)g=g-yp$,  equals the  expected payoff of the pure strategy No Crime, which is  $0$.  Setting $g -yp=0$  yields the equilibrium inspection rate $y^* = g/p$.
\end{itemize}

  Thus the unique MSNE is $(x^*,y^*) = (k/r,g/p)$.  This result demonstrates a key finding of the Inspection Game: a player's equilibrium randomization probability is determined entirely by the opponent's payoffs (e.g., $x^*$  depends on $k$ and $r$, which are inspector's payoffs). This is the ``paradoxical" effect that leads to the conclusion that the crime punishment $p$ has no effect on the chance of a citizen committing a crime \cite{Tsebelis_1989}. This paradoxical result provides the impetus for exploring the dynamics of the game using an evolutionary approach.
    The static solution to the Inspection Game, the unique MSNE, relies on the assumption of perfectly rational players who consciously randomize their actions to make their opponent indifferent.

  To address these limitations, we shift from classical game theory to evolutionary game theory \cite{Maynard_1982,Hofbauer_1998}.  Here, the MSNE probabilities are re-interpreted as population frequencies: $x$ is the proportion of citizens choosing Crime, and $y$ is the proportion of inspectors choosing Inspect. The replicator equation models a continuous dynamic process where strategies with higher-than-average payoffs increase their representation in the population \cite{Hofbauer_1998}. This dynamic approach does not assume rationality; instead, it models adaptation or learning.  In particular, we will show that the MSNE is not asymptotically stable under the replicator dynamics; rather, the system exhibits periodic oscillatory solutions,which better reflect the observed cyclical nature of crime and enforcement in real-world systems. However, writing down the replicator equations for the inspection game is not straightforward when the populations are asymmetric (i.e., the number of citizens $N $  is not equal to the number of inspectors $M$). Therefore, we first introduce the imitation dynamics. This approach allows us to simulate the game for finite $N $ and $M $ populations and provides a transparent micro-foundation for deriving the deterministic replicator equations when these populations are assumed to be infinite.

 \section{Imitation dynamics}\label{sec:id}

Consider two populations: $N$ citizens and $M$ inspectors.  Let $X$ be the number of citizens choosing Crime (the remaining $N-X$ choose No Crime), and let $Y$ be the number of inspectors choosing Inspect (the remaining $M-Y$  choose Not Inspect). The population frequencies of the strategies  Crime and Inspect are $x=X/N$ and $y=Y/M$, respectively.

 At each time step $\delta t$, a focal citizen $l_c$ and a focal  inspector $l_i$ are randomly chosen. They play a round of the inspection game with each other. They receive deterministic  payoffs $f_{l_c}$ and $f_{l_i}$, according to the payoff matrices (\ref{citizen}) and (\ref{inspector}) based on their chosen strategies.   Then,  a model citizen $m_c $ and a model inspector $m_i$,  different from $l_c$ and $l_i$,  are randomly chosen and similarly play a round of the game, resulting in payoffs  $f_{m_c}$ and $f_{m_i}$.  
 
 Focal individuals only update their strategies by imitating a more successful peer. Thus, $l_c$ and $l_i$ do not change their strategies if $f_{m_c} \leq f_{l_c}$ and $f_{m_i} \leq f_{l_i}$. However, when  $f_{m_c} > f_{l_c}$,  the  probability that the focal citizen $l_c$  switches to the strategy of the model citizen $m_c$  is proportional to the positive payoff difference
\begin{equation}\label{probc}
 \frac{f_{m_c} - f_{l_c}}{\Delta}.
\end{equation}
The parameter $\Delta$ is chosen so as to guarantee that the probability (\ref{probc})  is no greater than $1$.  
If citizens $l_c$ and $m_c$ have different strategies,  the  numerator is either $g$ (when a law-abiding citizen considers adopting the strategy of an uncaught criminal)  or $p-g$ (when a punished criminal considers switching to law-abiding behavior).

Similarly, when  $f_{m_i} > f_{l_i}$,  the  probability that the focal inspector $l_i$  switches to the strategy of the model inspector $m_i$  is
\begin{equation}\label{probi}
 \frac{f_{m_i} - f_{l_i}}{\Delta} .
\end{equation}
If the  inspectors $l_i$ and $m_i$ have different strategies,  then  the numerator of this equation can be either  $k$ (when a penalized inspector adopts the strategy of an inspector who chose not to inspect) or $r-k$  (when an inspector successfully adopts the strategy of a rewarded inspector).   

To ensure that the evolutionary rate reflects the relative intensity of selection across both populations, we choose  the  normalization factor
\begin{equation}\label{D}
\Delta = \max(g,p-g,k,r-k). 
\end{equation}
This choice couples the timescales of the two populations by normalizing the highest potential gain from switching strategies across the entire system.

Although citizens $l_c$ and  $m_c$ might use the same strategy, their payoffs can vary (e.g.,  $f_{m_c} > f_{l_c}$) because they interact with different opponents ($l_c$ with $l_i$, and $m_c$ with $m_i$).   In this scenario,  if $l_c$ were to imitate $m_c$, it would not alter the population composition.  

After the attempted strategy update, the time step $\delta t$ ends, and the time variable $t$ is updated to $t + \delta t$. The simulation continues until the stochastic dynamics converge to an absorbing state.  It is essential to note that in a finite population under these purely imitative dynamics, the system is a finite Markov process that must eventually converge to an absorbing state---that is, the fixation of one strategy in each population (e.g., all citizens choose Crime, $X=N$, or all choose No Crime,  $X=0$). The MSNE, which represents a stable mixed state of both strategies, is therefore never a long-term stable outcome in the finite stochastic model.

There are two primary purposes for using imitation dynamics.  First,  we use the dynamics in the limit of infinite populations ($N \to \infty$ and $M \to \infty$) and with infinitesimal time steps ($\delta t \to 0$). In this deterministic limit, the fixation boundaries cease to be absorbing states, and the system is accurately approximated by the  replicator differential equations. We will use these equations to analyze the stability and non-convergent periodic patterns centered on the MSNE. Second,  we use the stochastic dynamics to analyze the fixation probabilities of the different strategies in each finite population. 
The fixation analysis provides insights into the influence of stochasticity and population size on the long-term prevalence of strategies, even if the deterministic model predicts stable oscillations.

In the next section, we will derive the replicator differential equations by using the transition probabilities of the birth-death process corresponding to the imitation dynamics. We adopt this methodology because previous, standard methods used to derive the replicator equation from imitation dynamics---such as those based on the Fokker-Planck equation or general conditions for one-population or symmetric dynamics \cite{Traulsen_2005,Sandholm_2010,Fontanari_2024b}---do not generalize easily, or at all, to the asymmetric two-population scenario ($N \neq M$) with explicit timescale coupling required by the Inspection Game. Thus, to ensure a clear and robust micro-foundation for the asymmetric dynamics, we perform the derivation from first principles, establishing the necessary connection between the finite-population stochastic model and its deterministic limit.

 \section{Birth-death process}\label{sec:bd}

Here we use the framework of the standard birth-death process \cite{Karlin_1975,Antal_2006} to derive the dynamic equations. The state of the system is defined by the probability $P(X,Y;t)$ that at time $t$ there are $ X$ citizens choosing the strategy Crime and $Y$ inspectors choosing the strategy Inspect.

The core of this approach lies in the transition probabilities $T_{n,m}(X,Y)$, which represent the probability that the number of criminals $X$ increases by $n \in \{-1,0,1\}$ and the number of inspectors $Y$ increases by $m \in \{-1,0,1\}$ in a single time step $\delta t$. These transitions are determined by the rules of the imitation dynamics introduced in the previous section.

The deterministic replicator equations are found by calculating the expected change in the population numbers,  $\mathbb{E}(\Delta X)$  and $\mathbb{E}(\Delta Y)$.  For citizens, the expected change in the number of criminals in time $\delta t$  is
\begin{equation}
\mathbb{E}(\Delta X) = \sum_{n=-1}^1 n \sum_{m=-1}^1 T_{n,m} (X,Y) . 
\end{equation}
The quantity $ \mathbb{E}(\Delta X/N) /\delta t$  provides the discrete-time approximation of the time derivative $dx/dt$  in the deterministic limit where the population sizes tend to infinity  ($N \to \infty$ and $M \to \infty$)  and the time step tends to zero ($\delta t \to 0$).  In this limit, the population frequencies $x=X/N$ and $y=Y/M$ are approximated by continuous functions, which leads directly to the replicator  equations.  The full expressions for the necessary non-zero transition probabilities $T_{n,m} (X,Y)$, which are  used  next in  the derivation of the replicator equations,   are provided in Appendix \ref{appA}.  
 
 \subsection{Derivation of the replicator equations}
 
 The deterministic dynamic equations are derived by calculating the expected change in the number of individuals playing each strategy, $\mathbb{E}(\Delta X)$  and $\mathbb{E}(\Delta Y)$, and then taking the continuous limit.
 
The expected change in the number of criminals, $X$, in time step  $\delta t$ is  found by summing over all transitions that change $X$, i.e., $\mathbb{E}(\Delta X) = T_{1,0} + T_{1,-1} - T_{-1,0}$.  Taking the limits  $N \to \infty$ and $M \to \infty$ in   equations (\ref{t10}), (\ref{t-10}), and (\ref{t1-1}) and keeping the lowest order terms only yields
 \begin{equation}
\mathbb{E}(\Delta X) = \frac{1}{\Delta} x(1-x) \left (g - yp \right).
\end{equation}
Therefore 
 \begin{eqnarray}\label{repx}
\frac{dx}{dt}  & = &  \lim_{\delta t \to 0} \frac{1}{\delta t} \lim_{N,M \to \infty } \frac{\mathbb{E}(\Delta X)}{N}  \nonumber \\
& = & \frac{1}{\Delta} x(1-x) \left (g - yp \right)
\end{eqnarray}
provided that we set 
 \begin{equation}
\delta t = \frac{1}{N} .
\end{equation}
 The expected change in the number of inspecting individuals, $Y$, is $ \mathbb{E}(\Delta Y) = T_{0,1}  - T_{1,-1} - T_{0,-1}$ and following the previous proceeding we obtain 
 \begin{equation}
\mathbb{E}(\Delta Y) = \frac{1}{\Delta} y(1-y) \left (rx-k \right),
\end{equation}
which leads to
 \begin{eqnarray}\label{repy}
\frac{dy}{dt}  & = &  \lim_{\delta t \to 0} \frac{1}{\delta t} \lim_{N,M \to \infty } \frac{\mathbb{E}(\Delta Y)}{M}  \nonumber \\
& = & \frac{1}{\alpha \Delta} y(1-y) \left (rx-k \right), 
\end{eqnarray}
where  
 \begin{equation}
\alpha =  \frac{M}{N}
\end{equation}
 is the population size ratio, which is assumed to be finite and gives the number of inspectors per citizen.  The system of coupled differential equations (\ref{repx}) and (\ref{repy})  describes the continuous-time evolutionary dynamics of the Inspection Game, where $\Delta =\max(g,p-g,k,r-k) $ is the normalization constant that  couples the timescales of the two populations,  and $\alpha$ determines the relative evolutionary speed between the two populations.
 
 The derived replicator equations naturally partition the payoff parameters into two critical ratios that define the system's equilibrium state and policy impact. The ratio $k/r$  acts as the deterrence threshold, as it represents the minimum crime incidence required to make inspection profitable for the police. If the crime rate $x$ falls below this threshold ($x < k/r$), inspectors are deterred, their numbers fall, and the crime rate rises again. Similarly, the ratio $g/p$  is the inspection threshold, defining the minimum inspection frequency required to make law-abiding behavior profitable for citizens. If the inspection rate $y$ falls below this threshold ($y < g/p$), citizens are incentivized towards crime, and the inspection rate rises again. These two thresholds determine the neutral fixed point  around which the dynamic oscillations occur.

 It is worth noting that the Inspection Game has been previously studied  within  the framework of deterministic evolutionary game theory,  using  the standard, timescale uncoupled replicator equations for two populations of equal size \cite{Andreozzi_2002}. That analysis demonstrated the existence of stable limit cycles, a crucial finding for the deterministic dynamics. However, the average frequencies of crime and inspection over a cycle were found to coincide precisely with the  MSNE (or the neutral fixed-point) values. This result, therefore, reinforced the central paradox: the long-term average crime rate remained independent of the absolute penalty $p$ and gain $g$.  Our current derivation, which starts from the underlying stochastic process for asymmetric populations ($N \neq M$), yields a fundamentally different system of replicator equations,  Eqs.  (\ref{repx}) and (\ref{repy}). The resulting explicit timescale coupling between the citizen and inspector dynamics is essential for connecting the deterministic limit to the finite-population analysis, which is the necessary step for resolving the MSNE paradox.

 \section{Solutions of the replicator equations}\label{sec:re}
 
 The first important result concerning the replicator equations (\ref{repx}) and (\ref{repy}) is that the   equilibrium solutions do not depend on the normalization factor $\Delta$ or the population size ratio $\alpha$. However, these factors critically influence  the oscillatory solutions, which are the stable long-term outcome of the deterministic dynamics. For the sake of completeness, we first briefly discuss the equilibrium solutions and then characterize the oscillatory solutions.
 
 The equilibrium solutions  of the replicator equations  (\ref{repx}) and    (\ref{repy}), denoted by  $x^*$ and $y^*$, are obtained by setting $dx/dt=dy/dt=0$. Their  local stability is determined  by linearizing these equations at $x^*$ and $y^*$, resulting in the linear system 
\begin{equation}\label{linear}
\begin{pmatrix} 
du/dt \\ 
dv/dt
\end{pmatrix} = \mathbf{A}
\begin{pmatrix} 
u \\ 
 v   
 \end{pmatrix} ,
\end{equation}
where $u=x-x^*$, $v=y-y^*$,  and  $\mathbf{A}$  is the Jacobian matrix 
\begin{equation}
 \mathbf{A} = \frac{1}{\alpha \Delta  }
 \begin{pmatrix} 
 \alpha (1-2x^*)(g-y^*p) & -\alpha px^*(1-x^*)\\ 
r y^*(1-y^*) & (1-2y^*)(rx^*-k) 
\end{pmatrix} .
\end{equation}
The local stability of the equilibrium solutions is determined by the signs of the real parts of the eigenvalues of  $\mathbf{A}$  \cite{Britton_2003,Murray_2007}.  We briefly describe the  five equilibria below,  assuming the conditions for an interior mixed equilibrium are met ($p>g$ and $r>k$).

 \begin{enumerate}
 
 \item  Boundary Equilibrium: $x^* = 1$ and $ y^*=1$ (All Crime, All Inspecting).
The eigenvalues are $\lambda_c = (p-g)/\Delta $  and  $\lambda_i = -(r-k)/(\alpha \Delta )$.  Since $\lambda_c > 0$ and  $\lambda_i < 0$,  this equilibrium is a saddle point. This result reflects that while this   population cannot be invaded by inspectors who choose not to inspect,  it  is vulnerable to invasion by law-abiding citizens.

\item  Boundary Equilibrium: $x^* = 0$ and $y^*=1$ (All Law-Abiding, All Inspecting).  The eigenvalues are $\lambda_c = -(p -g)/\Delta$  and  $\lambda_i = k/(\alpha \Delta )$.  This is also a saddle point: the population cannot be invaded by criminals but can be invaded by inspectors who choose not to inspect.

\item  Boundary Equilibrium: $x^* = 1$ and $y^*=0$ (All Crime, All Not Inspecting). The eigenvalues are $\lambda_c = -g/\Delta $  and  $\lambda_i = (r-k)/(\alpha \Delta)$.  This is a saddle point: the population cannot be invaded by law-abiding citizens but can be invaded by inspectors who choose to inspect.

\item  Boundary Equilibrium: $x^* =0$ and $y^*=0$ (All Law-Abiding, All Not Inspecting).   The eigenvalues are $\lambda_c = g/\Delta $  and  $\lambda_i = -k/(\alpha \Delta )$.  This is a saddle point:  the population cannot be invaded by inspectors who choose to inspect  but can be invaded by criminals. 

\item  Interior Equilibrium:  $x^*= k/r$  and  $y^* = g/p$ (Mixed-Strategy Nash Equilibrium). This equilibrium corresponds to the coexistence of all four strategies.  The eigenvalues $\lambda_c$  and $\lambda_i$ are a conjugate pair of purely imaginary numbers.  The real part is zero, meaning this equilibrium is a neutral center. The imaginary part is given by
\begin{equation}
\mbox{Im} ( \lambda_c)  = \frac{1}{\Delta} \sqrt{ \frac{kg(p-g)(r-k)}{ \alpha r p}} .
\end{equation}

\end{enumerate}

Since the system has no stable equilibria, the solutions to the replicator equations (\ref{repx}) and (\ref{repy}) oscillate around the neutral fixed point $x^*= k/r$  and  $y^* = g/p$ \cite{Britton_2003,Murray_2007}.  The period of the oscillations of vanishingly small amplitude around this neutral fixed point is $T_{small}=2\pi/ \mbox{Im}(\lambda_c)$.

The phase plane  trajectories (or orbits) are the solutions of the single first-order differential equation obtained by factoring out time from the replicator system, 
\begin{equation}
\frac{dy}{dx}   = \frac{1}{\alpha}   \frac{y(1-y)(rx-k)}{x(1-x)  ( g-y p)}.
\end{equation}
This equation can be readily integrated by separation of variables to yield the closed-form constant of motion,  
\begin{equation}\label{orb1}
H = y^{\alpha g} (1-y)^{\alpha(p-g)}  x^{k}  (1-x)^{r-k} .
\end{equation}
The existence of this conserved quantity, which is often referred to as the Hamiltonian of the system,  indicates that the system is conservative. This means the phase plane trajectories are closed orbits surrounding the neutral center  $(x^*, y^*) =( k/r, g/p)$, and the exact orbit is uniquely determined by the initial conditions $x_0=x(0)$ and $y_0=y(0)$.

Note that the shape of these trajectories (level sets of $H$) does not depend on the normalization  $\Delta$, as   $\Delta$  only dictates the overall speed, or timescale, of the dynamics (and thus influences the period  $T$ of the oscillations) and disappears when time is factored out in the  phase plane analysis.  In contrast, the population size ratio $\alpha$ critically influences the geometry of the orbits. As seen in eq.  (\ref{orb1}), $\alpha$ acts strictly as a scale factor on the citizen's payoff parameters ($g$ and $p-g$) in the exponents of the $y$ terms. This scaling effectively weighs the citizen population's selective pressure relative to the inspector population's and governs the aspect ratio and skewness of the closed orbits. For simplicity,  henceforth we set the cost of inspection to $k=1$ without loss of generality. This  parameter choice means that the other payoff parameters ($p$, $g$  and $r$) are measured in units of the cost $k$ incurred by an inspector for performing an inspection.

\begin{figure}[th] 
\center
 \includegraphics[width=1\columnwidth]{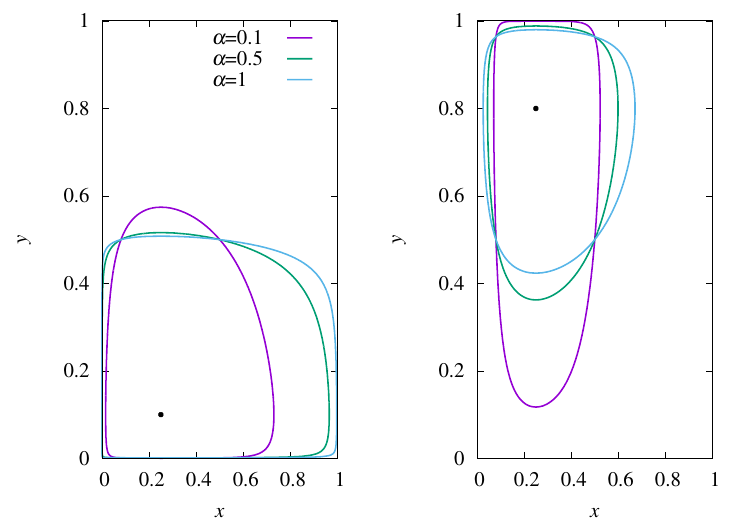}  
\caption{Phase plane trajectories showing  frequency  of criminals $x$ and frequency of inspectors who inspect  $y$.  The constant of motion $H$ defining each trajectory is set by the initial condition $x_0=y_0 = 0.5$.
The figure displays the influence of the population ratio  $\alpha $ on the orbit geometry for two different inspection thresholds $g/p=0.1$ (left panel) and $g/p=0.8$ (right panel).  In each panel, three trajectories for $\alpha = 0.1$, $0.5$, and $1$ are shown. The other parameters are fixed at
$g=4$,  $r=4$ and $k=1$.  Trajectories are counterclockwise and centered at the neutral fixed point $(x^*, y^*) =( k/r, g/p)$,  indicated as a filled circle.  
 }  
\label{fig:1}  
\end{figure}

The analytical findings of a neutral center and closed orbits are best visualized in the phase plane.  Figure \ref{fig:1}  illustrates trajectories for low ($g/p=0.1$) and high  ($g/p=0.8$) inspection thresholds, confirming the conservative nature of the system. The orbits are closed curves centered on the MSNE $(x^*, y^*) =( k/r, g/p)$, demonstrating the perpetual oscillation of the crime and inspection frequencies. The shape and aspect ratio of these closed orbits are directly influenced by the population size ratio,  $\alpha = M/N$.    The orbits exhibit a complex geometry determined by the non-linear structure of the Hamiltonian $H$ and the payoff parameters. Given the typical parameter ranges for the Inspection Game, the orbits are frequently asymmetric around the neutral fixed point, often displaying a skew toward the boundaries corresponding to higher payoff exponents.

For  high penalty $p$, the normalization constant $\Delta = p-g \approx p$ is large.  This results in a much slower rate of change for both frequencies, with the inspector population  changing very slowly with a rate proportional to $1/\Delta$ (see eq.  (\ref{repy})), leading to a long period for the entire cycle.   
 More importantly for the stochastic analysis, the counterclockwise trajectory passes very close to the boundary $x=0$ (see left panel of Fig. \ref{fig:1}).   This suggests that demographic noise is highly likely to lead to the extinction of criminals (fixation at $X=0$) in the high-penalty regime.
For light penalty  $p$, the normalization constant $\Delta = \max(g,k,r-k)$ is smaller and approximately constant, resulting in much faster dynamic.  More importantly,  the trajectory passes  close to the boundary $y=1$,  especially for small $\alpha$ (see the right panel of Fig. \ref{fig:1}).   The proximity to $y=1$ implies that demographic noise is likely to lead first to the fixation of inspectors who inspect (fixation at $Y=M$).  Once the inspector population  is fixed at $Y=M$, the deterministic flow dictates that the criminal frequency $x$ must decrease to zero,  resulting in the final absorbing state $(X=0,Y=M)$.  This is a surprising and key result for a scenario of light penalty, as it suggests that low penalties can paradoxically lead to the elimination of crime when finite population noise is present. We will return to this critical  issue in Section \ref{sec:fp}.

We note that if $T$ is the period of oscillations, we have from eq. (\ref{repy})
\begin{equation}
\frac{1}{T} \int_{y(0)}^{y(T)} \frac{dy}{y(1-y)} = r \frac{1}{T} \int_0^T dt x(t) - k
\end{equation}
and since the orbit is closed, $y(T)=y(0)$, the left-hand side and thus the right-hand side must vanish. Hence the average frequency of criminals over the oscillation period is
\begin{equation}
\frac{1}{T} \int_0^T dt x(t) = \frac{k}{r},
\end{equation}
which equals the neutral fixed point $x^*$. A similar analysis using eq. (\ref{repx}) leads to the conclusion that the average frequency of inspectors who inspect is $y^* = g/p$. Therefore, considering average time frequencies does not offer useful information to address the deterrence paradox, as the long-term crime frequency average remains fixed at the MSNE value, independent of the absolute penalty $p$.

The primary quantity of interest for policy analysis is the amplitude of the oscillations in the citizen population, specifically how the payoff parameters and the population size ratio $\alpha$  affect the minimum criminality incidence,  $x_{min}$,  and the maximum criminality incidence,  $x_{max}$.   Determining $x_{min}$ and $x_{max}$ involves solving the implicit equation for the Hamiltonian (\ref{orb1})  for $x$  subject to the condition that the inspector frequency is fixed at its neutral value  $y=y^*=g/p$.  The points on the orbit where $y=g/p$ correspond to the vertical tangency points of the orbit ($dy/dx \to \infty$),  which marks the maximum ($x_{max}$) and minimum ($x_{min}$) extent of the citizen oscillation.  Although  of not direct policy interest like the crime incidence amplitudes, the amplitudes $y_{max}$ and $y_{min}$  of the inspector population's oscillations are calculated similarly by setting $x=x^*=k/r$ in eq.  (\ref{orb1})  and are nevertheless crucial for understanding the effect of demographic noise in Section \ref{sec:fp}.

\begin{figure}[th] 
\center
 \includegraphics[width=1\columnwidth]{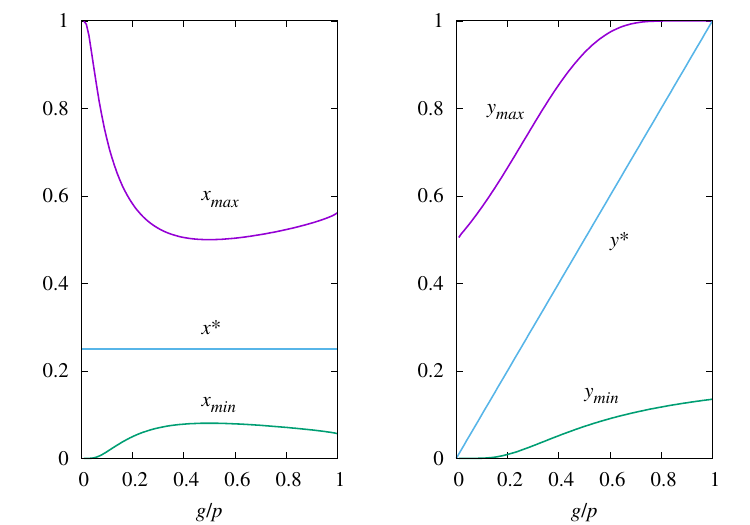}  
\caption{(Left panel) The minimum criminality incidence $x_{min}$,   the maximum criminality incidence $x_{max}$, and the neutral fixed point $x^*$.  (Right panel) The minimum inspector frequency $y_{min}$, the maximum inspector frequency $y_{max}$, and the neutral fixed point $y^*$. Both panels are shown as a function of  the inspection threshold $g/p$.  The initial condition is $x_0=y_0 = 0.5$.
The other parameters are fixed at $\alpha=0.1$,  $g=4$,  $r=4$, and $k=1$.   
 }  
\label{fig:2}  
\end{figure}

Figure \ref{fig:2} shows the effect of the inspection threshold $g/p$ on the amplitudes  of crime and inspection frequencies.  Note that,  for fixed $g$,  increasing the crime penalty $p$ decreases the inspection threshold. Interestingly, if the policy intention is to minimize $x_{max}$, then increasing $p$ is the worst possible action.  Paradoxically,  for large $p$,  the maximum criminality incidence approaches $x_{max} \approx 1$ (all citizens choose Crime), reaching an even higher peak than if the net penalty were negligible (i.e., $g/p \approx 1$).  This amplification of the crime cycle is a consequence of the inspection amplitude: $y_{max}$  reaches its minimum for large $p$ and its maximum for $p \approx g$. The maximum criminality incidence is minimized for $g/p = y_0$, where it takes the value $x_{max} = x_0$. At this point, the minimum criminality incidence is maximized, so the scenario corresponds to the minimum amplitude oscillation permissible for the fixed parameters and initial conditions.

\begin{figure}[th] 
\center
 \includegraphics[width=1\columnwidth]{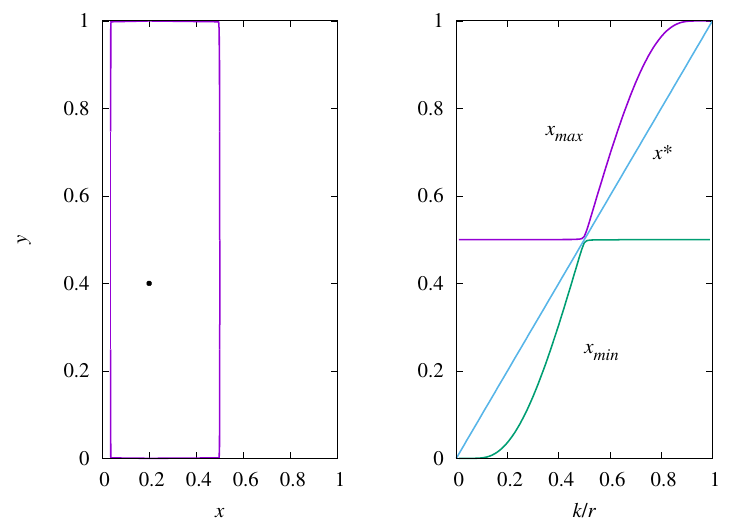}  
\caption{ (Left panel)  Phase plane trajectory showing criminal  frequency ($x$) and inspectors who inspect  frequency ($y$) for the population size ratio $\alpha =0.001$  and  $k/r=0.2$.  Trajectory  is counterclockwise and centered at the neutral fixed point $(0.2, 0.4)$,  indicated as a filled circle.  (Right panel) The minimum  ($x_{min}$) and  the maximum  ($x_{max}$)  criminality incidence, and the neutral fixed point ($x^*$) as a function of $k/r$ for  $\alpha =0.001$  The initial condition is $x_0=y_0= 0.5$. The other parameters are fixed at
 $p=10$, $g=4$, and $k=1$.  
 }  
\label{fig:3}  
\end{figure}

A particularly important limit is $\alpha \to 0$, meaning the number of inspectors ($M$) is much less than the number of citizens ($N$). Figure \ref{fig:3} shows a trajectory for a small population size ratio as well as the extremes of crime incidence frequency as a function of the ratio $k/r$. The orbits collapse onto a square-like limit cycle. The frequency of inspectors who inspect  ($y$) jumps instantly from $y=0$ to $y=1$ and back to $0$ since the right-hand side of eq.  (\ref{repy}) diverges. The dynamics of $x$ then slowly evolve along these boundaries following eq. (\ref{repx}) with $y$ set to $0$ or to $1$.

The limit $\alpha \to 0$ can be studied analytically.   The analytical boundaries of the relaxation cycle are defined by the points where the $\alpha$-dependent terms in the Hamiltonian (\ref{orb1}) balance out.  
For $y \approx 0$ (the boundary of the $x$-evolution) we can write the orbit equation in a form that highlights the limit,
\begin{equation}
y \approx  (\frac{x_0}{x})^{k/(\alpha g)} \left ( \frac{1-x_0}{1-x} \right )^{(r-k)/(\alpha g)}  y_0 (1-y_0)^{(p-g)/g}
\end{equation}
which makes it evident that $y \to 0$  provided that
\begin{equation}\label{cond}
\left ( \frac{x_0}{x} \right )^{k} \left ( \frac{1-x_0}{1-x} \right )^{r-k}  < 1. 
\end{equation}
Let us introduce the auxiliary function
\begin{equation}
h(x) = x^k (1-x)^{r-k} - x_0^k (1-x_0)^{r-k}
\end{equation}
such that $h(x) =0$  has exactly two roots, since the only root of $dh/dx = 0$ is $x = k/r$ (the coordinate of the neutral fixed point).  One root is obviously $x=x_0$.  The other root defines the extent of the oscillation, either $x_{min}$ or $x_{max}$,  and must be found numerically.  The satisfaction of condition (\ref{cond}) determines the interval of $x$ evolution:
\begin{itemize}
\item[-] In the case where the neutral fixed point $x^* = k/r$ is smaller than the initial condition ($k/r < x_0$), this other root yields $x_{min}$, and the condition (\ref{cond}) is satisfied for $x_{min} < x < x_0$ (this is the case shown in the left panel of Figure \ref{fig:3}).
\item[-] In the case where $k/r > x_0$, this root yields $x_{max}$, and condition (\ref{cond}) is satisfied for $x_0 < x < x_{max}$.
\end{itemize}

This analysis reveals that in the limit $\alpha \to 0$,  the  extremes $x_{max}$ and $x_{min}$ of the criminal  frequencies do not depend on the payoff parameters $p$ and $g$,  similarly to the coordinate of the neutral fixed point  $x^*$.    Due to this analytical finding, we choose to present the citizen oscillation amplitudes in the right panel of Fig.  \ref{fig:3}  as a function of the deterrence threshold $k/r$  rather than the inspection threshold $g/p$.

More importantly  for predicting the effect of the  demographic  noise,  the fixation path is determined by the relationship between the initial criminal frequency $x_0$ and the deterrence threshold $x^*=k/r$:
\begin{itemize}
\item[-] If $x_0 > k/r$: The inspector's expected payoff is positive, causing the fast $y$-dynamics to push $y$ toward $y=1$. Noise is therefore very likely to cause the first fixation at All Inspect ($Y=M$). Once the inspector population is fixed at $Y=M$, the subsequent deterministic flow drives the citizen population to criminals' extinction ($X=0$).
\item[-] If $x_0 < k/r$: The inspector's expected payoff is negative, causing the fast $y$-dynamics to push $y$ toward $y=0$. Noise is therefore likely to cause the first fixation at All Not Inspect ($Y=0$). Once the inspector population is fixed at $Y=0$, the subsequent deterministic flow drives the citizen population to the fixation of criminals ($X=N$).
\end{itemize}
This scenario will be corroborated in Section \ref{sec:fp} by the finite population simulations. Since the trajectories run counterclockwise, the dominance or demise of criminal behavior is determined entirely by whether the initial frequency of criminals is less or greater than the deterrence threshold $k/r$, with no influence from the citizen payoff parameters $g$ and $p$.

We can also obtain analytical results for the period of the oscillatory solutions in the singular limit $\alpha \to 0$. In this limit, the system undergoes a relaxation oscillation where the period $T$ is not zero but approaches a finite, non-zero value dominated by the slow evolution of the citizen population ($x$). This period is calculated by integrating the slow dynamics ($dx/dt$) along the boundaries $y=0$ and $y=1$.

The slow dynamics along the $y=0$ boundary is given by
\begin{equation}
\frac{dx}{dt} =  \frac{g}{\Delta} x(1-x) 
\end{equation}
and the time taken for $x$ to evolve from $x_{min}$ to $x_{max}$ (following the $\frac{dx}{dt} > 0$ flow; recall the trajectories run counterclockwise) is
\begin{equation}
T_0 = \int_{x_{min}}^{x_{max}} \left ( \frac{dx}{dt} \right )^{-1} dx = \frac{\Delta}{g} \int_{x_{min}}^{x_{max}} \frac{1}{x(1-x)} dx = \frac{\Delta}{g} \ln\left( \frac{x_{max}(1-x_{min})}{x_{min}(1-x_{max})} \right) .
\end{equation}
Similarly, the time taken for $x$ to evolve from $x_{max}$ to $x_{min}$ along the $y=1 $ boundary  is 
\begin{equation}
T_1 = \frac{\Delta}{p-g} \ln\left( \frac{x_{max}(1-x_{min})}{x_{min}(1-x_{max})} \right) .
\end{equation}
So the final analytical expression for the total period of the relaxation oscillation $T  = T_0+T_1$  is
 \begin{equation} \label{Ta0}
 T =  \frac{p \Delta}{g(p-g)} \ln\left( \frac{x_{max}(1-x_{min})}{x_{min}(1-x_{max})} \right) .
 \end{equation} 
The period vanishes when $x_{min} = x_{max} = x_0 = x^* = k/r $,  corresponding to a vanishingly small amplitude oscillation for the slow variable $x$. We note, however, that in the $\alpha=0$ limit,   we have $y_{max}-y_{min} = 1$,  so the small amplitude scenario never hold for the fast variable $y$.   

For the case of light penalty $g/p \approx 1$ we have $\Delta = \max(g,k,r-k)$.   Substituting $\Delta$ into the period equation $T$, the term $p/(p-g)$ diverges as $p \to g$. This divergence confirms that the period becomes arbitrarily long as the system approaches the neutral boundary fixed points, preventing the relaxation cycle from forming. For high penalty $g/p \ll1$, we  use $\Delta = p-g$. Substituting this into eq. (\ref{Ta0})  yields  $T \propto p/g$, since $x_{min}$ and $x_{max}$ depend only on the initial condition $x_0$ and the ratio $k/r$.

This result demonstrates that as the inspector population becomes infinitesimally small ($\alpha \to 0$), the oscillation's timescale is determined entirely by the slow evolution of the citizen population. This is because the period is dominated by the time required for citizens to react to the extremes of inspection ($y=1$) and non-inspection ($y=0$), indicating that the limit cycle behavior is a slow-fast phenomenon with a characteristic time scale independent of the small parameter $\alpha$.

\section{Finite population analysis}\label{sec:fp}

Although the replicator equations predict intuitive oscillatory behavior, driven by the interplay of dominant and counter-strategies, demographic noise (arising from finite population size) destabilizes these solutions.  The noise drives the system towards absorbing states, which correspond to the boundary equilibria of the replicator equations.  Here we present Monte Carlo simulations  of the imitation dynamics  for finite populations described in Section \ref{sec:id}. 

Let $\rho_{IJ} $ with $I \in \{0,N\}$ and $J \in \{0,M\}$ represent the probabilities of fixation for the four possible absorbing states: 
\begin{enumerate}
\item ($I=N$, $J=M$): All citizens commit crime, all inspectors inspect.
\item ($I=0$, $J=M$): All citizens are law-abiding, all inspectors inspect.
\item ($I=N$, $J=0$): All citizens commit crime, all inspectors do not inspect.
\item ($I=0$, $J=0$): All citizens are law-abiding, all inspectors do not inspect.
\end{enumerate}
The sum of these probabilities is unity,
  $\rho_{NM}+\rho_{0M}+\rho_{N0} + \rho_{00} = 1$.  These probabilities are estimated empirically from $10^5$ independent stochastic simulations for each parameter configuration.  To facilitate visualization, here we focus only  on the probability that criminality is  extinct, i.e., $\rho_0 =  \rho_{0M}  + \rho_{00}$. 
  
\begin{figure}[th] 
\center
 \includegraphics[width=1\columnwidth]{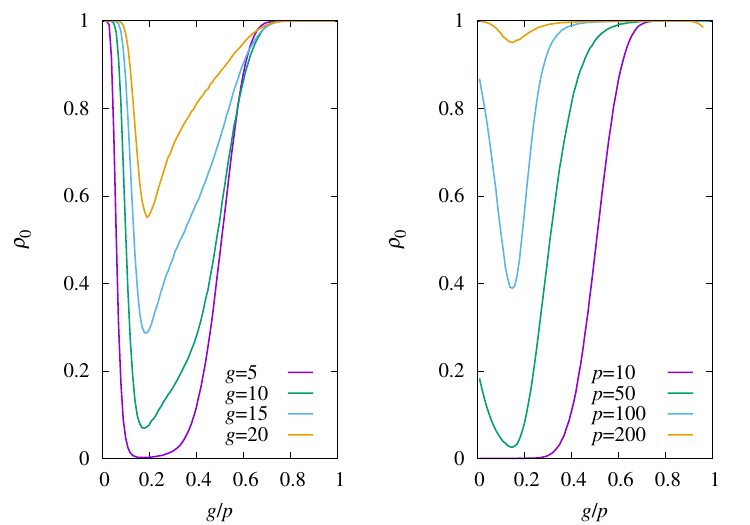}  
\caption{ Probability that criminality is extinct  $\rho_0$ as a function of the inspection threshold $g/p$ for a citizen population size of $N=1000$ and an inspector population size of $M=100$ ($\alpha=0.1$).   (Left panel) Varies the penalty parameter $p$  for different fixed values of the crime gain $g=5,10,15,20$.  (Right panel) Varies the crime gain $g$ for different fixed values of  the penalty $p=10,50,100, 200$.
The initial condition is $x_0=y_0 = 0.5$.
The other parameters are fixed at  $r=4$ and $k=1$.  
 }  
\label{fig:4}  
\end{figure}

Figure \ref{fig:4}   shows the probability that a citizen population is  free from criminals ($\rho_0$) in the long run as a  function of the inspection threshold $g/p$ for a system with a small inspector population ($\alpha =0.1$).  The two panels illustrate that the influence of $p$ and $g$ on fixation is not determined only by their ratio.  The  left panel, was  obtained by fixing  $g$ and varying $p$ in the range $(g, \infty)$, while the right panel was obtained by fixing $p$ and varying $g$ in the range $(0,p)$. 

For fixed $g$, the left panel of Fig. \ref{fig:4} shows that there are two distinct  regimes where crime can be almost completely eradicated: for very large $p$ (i.e.,  $g/p \to 0$)  and for $ p \approx g$ (i.e., $g/p \to 1$).  In the  $g/p \to 0$ regime,  the probability of strategy switching for inspectors is vanishingly small ($\propto 1/\Delta$  with $\Delta \approx p \gg 1$),   while criminals are virtually guaranteed to switch to law-abiding behavior when caught.  
This high probability of criminal extinction agrees with our analysis of the effect of stochastic noise on the phase plane trajectory shown in the left panel of Figure \ref{fig:1}, which passes very close to the $x=0$  boundary. 
This result restores the relevance of high penalties for suppressing crime under noise, challenging the deterministic result. However, the probability of crime fixation
($1-\rho_0$) increases sharply as $p$ decreases, and then, surprisingly, begins to decrease again, disappearing altogether in the $p \approx g$ regime. This fixation process, while exceedingly slow, can be inferred from the phase plane (right panel of Figure \ref{fig:1}): noise drives the system to the fixation of inspectors ($Y=M$), which then dooms the criminals to extinction ($X=0$).

The right panel of Fig. \ref{fig:4}, obtained by fixing $p$ and varying $g$, shows that the probability of crime extinction remains high for low $g/p$ (high penalty) only provided that the penalty $p$ is sufficiently large in absolute terms. This is because the mechanism for crime extinction in this regime relies on the slow evolution of the inspector population ($y$), which requires the normalization factor $\Delta = p-g$ to be large. In other words, $g/p$ can be small because $g$ is small, but if $p$ is also small (e.g., $p=10$), this does not result in effective crime suppression. On the other hand, crime is robustly suppressed in the regime $g/p \approx 1$ across all penalty values, further demonstrating the surprising finding that light penalties can consistently suppress crime in a noisy environment.

\begin{figure}[th] 
\center
 \includegraphics[width=1\columnwidth]{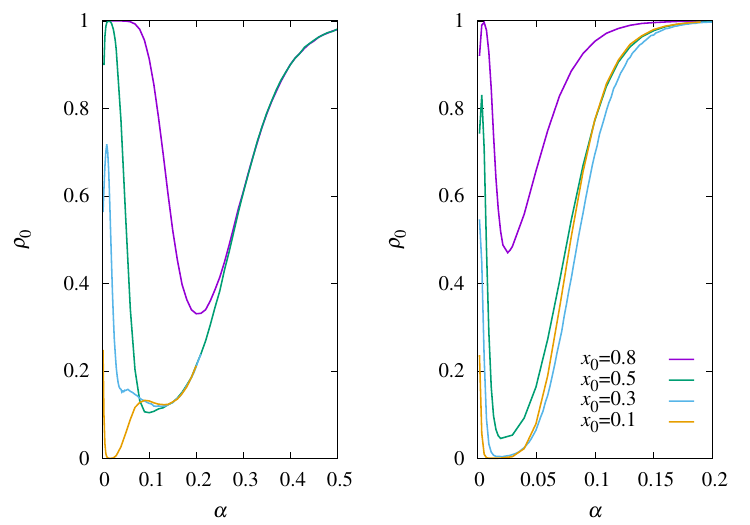}  
\caption{Probability that criminality is extinct  $\rho_0$ as a function of the population size ratio $\alpha$ for a citizen population size of $N=1000$ and different proportion of criminals in the initial population ($x_0=0.1,0.3,0.5,0.8$).   (Left panel) Penalty parameter $p=10$.    (Right panel) Penalty parameter $p=100$.   The initial condition for the inspector population is $y_0 = 0.5$.
The other parameters are fixed at  $g=4$, $r=4$ and $k=1$.  
 }  
\label{fig:5}  
\end{figure}

To conclude our analysis, in Fig.  \ref{fig:5} we present the effects of the population size ratio $\alpha$ and the initial proportion of criminals $x_0$ on the probability that crime is suppressed ($\rho_0$). We find a complex interplay between these parameters only for small $\alpha$. Otherwise, the initial fraction of criminals has practically no effect on $\rho_0$, which increases monotonically with increasing $\alpha$.

The scenario is much more complex for  small $\alpha$. As predicted by our analytical study of the limit $\alpha=0$,  the fixation or the extinction of the criminals depends on whether  their initial fraction is less than or greater than $x^*=k/r$.  The results of Fig. \ref{fig:5} for finite populations indicate that  criminal dominance ($\rho_0$ is small )  occurs when the initial criminal frequency is  small.  This  result is related to the fact that in asymmetric games the slowest and initially rarest strategy is likely to prevail \cite{Vieira_2025}.  The logic is simple. For small $\alpha$,   the evolution of the inspector population is much faster than that of the citizen population.   In the deterministic limit, this is evident in  the factor $\alpha$  that appears in the replicator equation (\ref{repy}) for the frequency $y$,  and in the stochastic scenario,   this is due to the smallness of $M$, since the fixation time scales with the population size.   If there are very few criminals at the beginning,  the inspector population quickly fixates at  $Y=0$ (Not Inspect),  since this is the strategy  with the largest average payoff in a population with few criminals.  
Once inspectors fixate at $Y=0$, the guaranteed impunity drives the slow-evolving citizen population toward the fixation of All Crime ($X=N$).
We note that the smallest values of $\alpha$ in the figure correspond to $M=2$ (since $N=1000$), where the stochastic effects are too strong to permit a clean mechanistic interpretation. 
The complex interplay of $\rho_0$ on $\alpha$ and $x_0$ revealed in Fig.  \ref{fig:5} is a consequence of the fact that the selective advantage of the rarer strategy diminishes as $\alpha$ increases, making the time scales of the inspector and citizen dynamics less dissimilar.

\section{Discussion}\label{sec:disc}

Our study of the Inspector Game using evolutionary game theory, particularly within the framework of finite, asymmetric populations, resolves a significant paradox left open by classical game theory. The traditional Mixed-Strategy Nash Equilibrium predicts that penalty size is irrelevant to long-term crime rates, suggesting that high penalties are not an effective deterrent. In contrast, our finite population analysis reveals that when demographic noise is included, high absolute penalty $p$ ($g/p \to 0$) is indeed a successful mechanism for suppressing crime, leading to the extinction of criminals. This fixation results from the large magnitude of $p$ slowing the inspector dynamics (via the large normalization factor $\Delta$), allowing the citizen population to quickly fixate on law-abiding behavior. Furthermore, a robust and counter-intuitive finding holds across both the infinite-population deterministic and the finite-population stochastic models: light penalty ($g/p \approx 1$) also effectively suppresses crime. This unexpected stability underscores the complex and nuanced relationship between deterrence parameters and long-term behavioral outcomes in dynamic, evolving populations.

Our analysis of the $\alpha \to 0$ limit---where the inspector population ($M$) is infinitesimally small compared to the citizen population ($N$)---provides crucial insights applicable to realistic resource constraints. In this highly asymmetric regime, the system exhibits slow-fast relaxation dynamics, with the inspector frequency ($y$) reacting nearly instantaneously, while the criminal frequency ($x$) evolves slowly along the boundaries.  Crucially, this massive decoupling in timescales in the asymptotic limit $\alpha \to 0$ supersedes the general deterrent effect of the penalty $p$ found in the finite and moderately asymmetric cases. Most importantly, we demonstrated that the ultimate fate of the system (criminal dominance or demise) is determined entirely by the initial criminal frequency ($x_0$) relative to the deterministic deterrence threshold $x^*=k/r$. If the initial crime rate is above $k/r$, the fast dynamics drive the system toward inspection fixation, leading to criminal extinction. Conversely, if $x_0$ is below $k/r$, the system is driven toward non-inspection, guaranteeing criminal dominance. This result indicates that in resource-scarce environments, the initial state of the population, set against the cost/reward ratio $k/r$, acts as the primary predictor of long-term crime rates,  rendering  the citizen payoff parameters ($p$ and $g$) irrelevant in this specific asymptotic regime.

Surprisingly, the asymptotic result in the $\alpha \to 0$ limit can be seen as partially vindicating the structure of the classical MSNE conclusion. By showing that the long-term outcome is independent of the citizen payoff parameters $p$ and $g$, our  analysis reproduces the core formal characteristic of the static game's solution. However, this vindication comes with a new layer of complexity: unlike the MSNE, which predicts a stable non-fixated frequency, our dynamic prediction results in fixation (either All Crime or Extinction) dictated solely by the initial condition ($x_0$) and the inspection parameters ($k$ and $r$).  
Given that the classical model, by design, cannot account for population size or initial state, the  independence from $p$ and $g$  in the $\alpha \to 0$ limit revealed here could be regarded as a novel paradox of crime enforcement theory.

The robust finding across all simulation contexts is the paradoxical relationship between the inspection threshold ($g/p$) and the long-term success of crime suppression ($\rho_0$). Our finite population simulations reveal a U-shaped policy outcome: crime is highly likely to be eliminated when the threshold is approached from either extreme---very low ($g/p \to 0$, high absolute penalty) or very high ($g/p \approx 1$, light penalty). In the former case, success relies on the magnitude of $p$ slowing the system; in the latter, success hinges on the initial fast dynamics leading to inspection fixation ($Y=M$). The greatest risk of criminal dominance ($1-\rho_0 \approx 1$) occurs in the intermediate regime where the penalty is moderate. This analysis suggests that policies should aim for the extremes, either implementing strong, absolute deterrence or relying on the robust, self-correcting dynamics found under light penalties.

Finally, our work demonstrates that the complex dynamics of the Inspection Game, which evade simple prediction under static equilibrium analysis, are only fully captured by integrating evolutionary dynamics and demographic noise. Moving beyond the limitations of the Mixed-Strategy Nash Equilibrium, we show that the long-term success of crime suppression is not a simple function of penalty ratio but depends critically on absolute penalty magnitude ($p$) and the initial conditions ($x_0$) relative to the deterrence threshold $k/r$ in the realistic limit of very few inspectors per citizen  ($\alpha \to 0$). Future research should build on this foundation by exploring models with heterogeneous payoff parameters \cite{Santos_2006,Perc_2015}, co-evolving population sizes (variable $\alpha$) \cite{Hansen_2024}, or complex social network structures \cite{Szabo_2007} to better approximate real-world enforcement scenarios. Ultimately, our findings underscore a core message for policy: effective crime strategy cannot rely solely on rational expectations but must instead be informed by the dynamic, stochastic nature of evolving human populations.

\backmatter

\bmhead{Acknowledgments}

JFF is partially supported by  Conselho Nacional de Desenvolvimento Cient\'{\i}fico e Tecnol\'ogico  grant number 305620/2021-5.

\begin{appendices}

\section{Transition Probabilities}\label{appA}

\renewcommand{\thefigure}{A\arabic{figure}}
\setcounter{figure}{0}

The full expressions for the single-step transition probabilities $T_{n,m}(X,Y)$ for the imitation dynamics in the asymmetric two-population Inspection Game are derived below. These probabilities are used to calculate the expected change in the number of individuals playing each strategy, $\mathbb{E}(\Delta X)$ and $\mathbb{E}(\Delta Y)$, which form the basis for the continuous-time replicator equations.

\subsection{Change in citizens only ($T_{1,0}$ and $T_{-1,0}$)}

The number of criminals, $X$, increases by one ($T_{1,0}$) only if a law-abiding focal citizen switches to Crime. The calculation involves considering the selection probabilities of the four players (focal/model citizen, focal/model inspector) and the successful imitation probability   $\Delta f/\Delta$, 
 \begin{eqnarray}\label{t10}
 T_
{1,0}(X,Y) & = & \left ( \frac{N-X}{N} \frac{M-Y}{M}\right ) \left ( \frac{X}{N-1} \frac{M-Y-1}{M-1} \right )  \frac{g}{\Delta}  \nonumber \\
 &  & + \left ( \frac{N-X}{N} \frac{Y}{M}\right ) \left ( \frac{X}{N-1} \frac{M-Y}{M-1} \right )  \frac{g}{\Delta} (1 - \frac{k}{\Delta}) .
 \end{eqnarray}
The number of criminals decreases by one ($T_{-1,0}$) only if a criminal focal citizen switches to No crime,
 \begin{eqnarray}\label{t-10}
  T_{-1,0}(X,Y) & = & \left ( \frac{X}{N} \frac{Y}{M}\right ) \left ( \frac{N-X}{N-1} \frac{Y-1}{M-1} \right )  \frac{p-g}{\Delta} \nonumber \\
 &  &  +\left ( \frac{X}{N} \frac{Y}{M}\right ) \left ( \frac{N-X}{N-1} \frac{M-Y}{M-1} \right )  \frac{p-g}{\Delta}  .
  \end{eqnarray}

\subsection{Change in inspectors only ($T_{0,1}$ and $T_{0,-1}$)}

The number of inspecting individuals, $Y$, increases by one ($T_{0,1}$) only if an ineffective focal inspector switches to Inspect,
\begin{eqnarray}\label{t01}
 T_{0,1}(X,Y) & = &  \left ( \frac{X}{N} \frac{M-Y}{M}\right ) \left ( \frac{X-1}{N-1} \frac{Y}{M-1} \right )  \frac{r-k}{\Delta}  \nonumber \\
  &  &  + \left ( \frac{N-X}{N} \frac{M-Y}{M}\right ) \left ( \frac{X}{N-1} \frac{Y}{M-1} \right ) \frac{r-k}{\Delta} .
  \end{eqnarray}
The number of inspecting individuals decreases by one ($T_{0,-1}$) only if a inspecting focal inspector switches to Not inspect,
 \begin{eqnarray}\label{t0-1}
T_{0,-1}(X,Y) & = &  \left ( \frac{N-X}{N} \frac{Y}{M}\right ) \left ( \frac{N-X-1}{N-1} \frac{M-Y}{M-1} \right )  \frac{k}{\Delta} \nonumber \\
&  & + \left ( \frac{N-X}{N} \frac{Y}{M}\right ) \left ( \frac{X}{N-1} \frac{M-Y}{M-1} \right )  \frac{k}{\Delta} (1- \frac{g}{\Delta}) .
  \end{eqnarray}

 \subsection{Simultaneous Change ($T_{1,-1}$)}
 
 The only possible simultaneous change is $X$ increasing and $Y$ decreasing ($T_{1,-1}$), which occurs when a law-abiding citizen switches to Crime and an inspecting inspector switches to Not Inspect,
 \begin{equation}\label{t1-1}
 T_{1,-1}(X,Y)= \left ( \frac{N-X}{N} \frac{Y}{M}\right ) \left ( \frac{X}{N-1} \frac{M-Y}{M-1} \right )  \frac{g}{\Delta} \frac{k}{\Delta}.
  \end{equation}
 The other simultaneous transition probabilities, $T_{1,1}$,  $T_{-1,-1}$, and $T_{-1,1}$, are zero.

\end{appendices}

\end{document}